\documentclass[twocolumn,prb]{revtex4}
\usepackage{amsmath}
\usepackage{amsfonts}
\usepackage{amssymb}
\usepackage{graphics}
\usepackage{float}
\usepackage{graphicx}
\usepackage{epstopdf}
\usepackage[compact]{titlesec}
\usepackage{natbib}
\usepackage{threeparttable}

\begin{document}
\title{Scattering times and mobility with localized impurities in topological insulator films }
\author{Parijat Sengupta}
\email{parijats@bu.edu}
\author{Enrico Bellotti}
\affiliation{Photonics Center \\
Department of Electrical and Computer Engineering, Boston University, Boston, MA 02215}

\begin{abstract}
The zero gap surface states of a 3D-topological insulator host highly mobile Dirac fermions with spin locked to the momentum. The high mobility attributed to absence of back scattering is reduced in presence of impurities on the surface. In particular, we discuss and compare scattering times for localized impurities on the surface, scattering between states of opposite helicity located on different surfaces coupled through a hybridization potential, and the role of magnetic impurities. Magnetic impurities give rise to an additional spin suppression factor. The role of warped bands and its influence on topological factors that can enhance the overall surface mobility is examined.
\end{abstract}
\maketitle

\vspace{0.25cm}
\section{Introduction}
\vspace{0.25cm}
The zero gap helical surface states~\cite{qi2011topological} of a 3D topological insulator (TI) film can be split by an external magnetic field or a reduction of film thickness.~\cite{liu2010oscillatory} Such a band gap open topological insulator loses protection against back scattering~\cite{roushan2009topological,zhou2009theory} and has a reduced surface mobility. The suppression of mobility is also observed through large-angle scattering in presence of impurities on the surface of a TI with a consequent degradation of its charge transport properties. Further, a sufficiently thin film, with preserved time reversal symmetry, allows the possibility of a surface electron(a Dirac fermion) to back scatter~\cite{yin2014coulomb} by tunneling through the bulk and eventually occupy a state of opposite helicity on the bottom surface thus providing a pathway for velocity reduction. Inter and intraband scattering events on the Dirac hyperbolas~\cite{shan2010effective} of each surface, created by splitting the TI states with an external magnetic field or a layered ferromagnet offers a way to probe phenomena such as weak anti-localization,~\cite{bao2012weak,kim2011thickness} electric conductivity, and localized magnetic moments through magnetoresistance measurements. 

In this work, within a Boltzmann approximation, the impurity scattering time for surface carriers is evaluated (Sec. II) in a topological insulator film. The impurities are assumed to be non-interacting. Scattering times are evaluated for four cases 1)a zero-gap non-trivial topological insulator 2) in presence of inter-surface coupling 3) with substrate induced intrinsic dipole moments and 4) with a finite magnetic moment with an out-of-plane component that introduces a spin induced suppression factor. The corresponding semi-classical mobilities (Sec. III) derived from a linearized Boltzmann equation and a relaxation time approximation are discussed. The role of warping~\cite{xu2011topological} and its influence on overall transport behaviour is also highlighted. Results are collected in Sec. IV followed by concluding remarks. 

\vspace{0.25cm}
\section{Scattering rates for a localized impurity}
\vspace{0.25cm}
Surface states in topological insulators such as Bi$_{2}$Te$_{3}$, Bi$_{2}$Se$_{3}$, and Sb$_{2}$Te$_{3}$ are characterized by a linear dispersion for small values of the momentum vector and a single Dirac cone at the $ \Gamma $ point. The calculations carried out in this work are partly performed using a low-energy continuum four-band k.p Hamiltonian for 3D TIs. The four-band k.p Hamiltonian~\cite{liu2010model} is further simplified in to a two-band Dirac Hamiltonian that represents only the linearly dispersing surface states; additional modifications that arise on account of higher warping terms are included later for comparison. The four-band k.p Hamiltonian in the basis set of the four lowest low-lying states $ \vert P1_{z}^{+} \uparrow \rangle $, $ \vert P2_{z}^{-} \uparrow \rangle $, $ \vert P1_{z}^{+} \downarrow \rangle $, and $ \vert P2_{z}^{-} \downarrow \rangle $ is 

\begin{align}
\label{eqn1}
H(k) = \epsilon(k) + \begin{pmatrix}
M(k) & A_{1}k_{z} & 0 & A_{2}k_{-} \\
A_{1}k_{z} & -M(k) & A_{2}k_{-} & 0 \\
0 & A_{2}k_{+} & M(k) & -A_{1}k_{z} \\
A_{2}k_{+} & 0 & -A_{1}k_{z} & -M(k) \\
\end{pmatrix}
\end{align}
where $ \epsilon(k) = C + D_{1}k_{z}^{2} + D_{2}k_{\perp}^{2}$, $ M(k) = M_{0} + B_{1}k_{z}^{2} + B_{2}k_{\perp}^{2}$ and $ k_{\pm} = k_{x} \pm ik_{y}$. For Bi$_{2}$Te$_{3}$ and Bi$_{2}$Se$_{3}$, the relevant parameters are taken from Ref.~\onlinecite{zhang2009topological}. The Hamiltonian is diagonalized numerically to obtain eigen values. 

\noindent The simplified two-dimensional Dirac Hamiltonian for the surface states can be written as~\cite{zhang2009topological}
\begin{equation}
H_{surf.states} = \hbar v_{f}(\sigma_{x}k_{y} - \sigma_{y}k_{x})
\label{dss}
\end{equation}
where $ v_{f}$ denotes \textit{Fermi}-velocity and $\sigma_{i};{i = x,y} $ are the Pauli matrices. The corresponding density of states in standard notation is
\begin{subequations}
\begin{flalign}
D(\epsilon)= \int \dfrac{dk}{\left(2\pi\right)^{2}}\delta\left(\epsilon - E(k)\right) 
\label{simdos1}
\end{flalign}
Changing to polar coordinates,
\begin{flalign}
D(\epsilon) &= \dfrac{1}{\left(2\pi\right)^{2}}\int_{0}^{2\pi} d\theta \int_{0}^{\infty}k\,dk\dfrac{1}{\hbar v_{f}}\delta\left(k - \dfrac{\epsilon}{\hbar v_{f}}\right)  \notag \\
&= \dfrac{2}{2\pi\left(\hbar v_{f}\right)^{2}}\vert\epsilon\vert
\label{simdos2}
\end{flalign}
\end{subequations}
\noindent The density of states of electrons described by the two-band Hamiltonian (Eq.~\ref{dss}) with spin degeneracy included is linear in energy and vanishes at the Dirac point. In deriving this expression, we have used the property $ \delta\left(-x\right)  = \delta\left(x\right) $.  

This form of the Hamiltonian, though in principle sufficient to probe the surface states, fails to account for the underlying crystal symmetries.~\cite{alpichshev2010stm,souma2011direct,wang2011observation} A more accurate representation (Eq.~\ref{warping}) within the two-band model~\cite{fu2009hexagonal} that is consistent with the C$_{3v}$ point-group symmetry of the rhombohedral crystal, the snow-flake Fermi contour obtained from ARPES, and also preserves time reversal symmetry must contain higher order terms cubic in $ k $. The next set of corrections involve $ k^{5} $ terms~\cite{basak2011spin} which are ignored in all calculations here. 
\begin{equation}
H(k) = \epsilon_{0}(k) + \hbar v_{f}(\sigma_{x}k_{y} - \sigma_{y}k_{x}) + \dfrac{\lambda}{2}\hbar^{3}\left(k_{+}^{3} + k_{-}^{3}\right)\sigma_{z}
\label{warping}
\end{equation} 
$ \epsilon_{0}(k) $ introduces the particle-hole anisotropy and the cubic terms denote warping. Using Eq.~\ref{warping}, and ignoring particle-hole anisotropy without loss of generality, the surface state spectrum is
\begin{equation}
\epsilon_{\pm}(k) =  \pm \sqrt{\hbar v_{f}^{2}k^{2} + \lambda^{2}\hbar^{6}k^{6}cos^{2}(3\theta)}
\label{warpsp} 
\end{equation}
\noindent where $ \theta = tan^{-1}\left(k_{y}/k_{x}\right) $. 

\noindent The scattering rate on surface of a topological insulator due to localized and static non-magnetic impurity with a spherical potential $ V = V_{0}\delta\left(r - R_{j}\right) $ can be evaluated using the Fermi-Golden rule. The concentration of impurities is assumed to be dilute such that there are no interference effects between successive scattering events. The matrix element is
\begin{subequations}
\begin{equation}
M\left(k^{'},k\right) = \langle\Psi_{f}\vert V \vert\Psi_{i}\rangle 
\end{equation}
where the two-component wave function $ \Psi $ for the linear Hamiltonian (Eq.~\ref{dss}) is 
\begin{equation}
\Psi_{\eta} = \dfrac{1}{\sqrt{2}}\begin{pmatrix}
\lambda_{\eta}(k)exp(-i\theta) \\
\eta \lambda_{-\eta}(k)
\end{pmatrix}
\label{wfun1}
\end{equation}
and 
\begin{equation}
\lambda_{\eta}(k) = \sqrt{1 \pm \dfrac{\Delta}{\sqrt{\Delta^{2}+ \left( \hbar v_{f}k\right)^{2}}}}
\label{wfun2}
\end{equation}
\end{subequations}

\vspace{0.25cm}
\subsection{Scattering rate between states of identical helicity}
\vspace{0.25cm}
The helicity of the electron wave function is denoted by $ \eta = \pm $ and $ \Delta $ represents a band gap for the surface states. Evaluating the square of the matrix element for a ungapped TI film when an electron elastically scatters through an angle $ \phi $ between states with identical helicity gives 
\begin{eqnarray}
\vert M\left(k^{'},k\right)\vert^{2} &=& \vert\langle\Psi_{f}\vert V_{0} \vert\Psi_{i}\rangle\vert^{2} 
= V_{0}^{2}cos^{2}\dfrac{\phi}{2}
\label{matlemsq}
\end{eqnarray}
and $ \phi = \theta_{f}-\theta_{i} $.
The elastic scattering time using a Boltzmann approximation~\cite{mahan2000many,han2013modern} such that $ \vert k \vert = \vert k^{'} \vert $ is
\begin{equation}
\dfrac{1}{\tau} = \dfrac{2 \pi}{\hbar}\int \dfrac{d^{3}k^{'}}{8\pi^{3}}\delta\left(\varepsilon_{k} - \varepsilon_{k^{'}}\right)\vert \chi_{kk^{'}}\vert\left(1 - cos\phi\right)
\label{simbz}
\end{equation}
where $ \chi_{kk^{'}} = \vert M\left(k^{'},k\right)\vert^{2}\zeta\left(s,s^{'}\right) $. The additional spin-scattering factor $ \zeta\left(s,s^{'}\right) $ takes in to account the helical spin structure of the TI surface states. For a pristine TI, $ \zeta\left(s,s^{'}\right) = cos^{2}\dfrac{\phi}{2} $ with $ 0 \leqslant\phi\leqslant \pi $, since back scattering is forbidden and the factor must assume a value between zero and unity. The integral in Eq.~\ref{simbz} therefore must be integrated over all values of $ \phi $.
\begin{flalign}
\dfrac{1}{\tau} &= \dfrac{2 \pi}{\hbar}g\left(\varepsilon\right)\int_{0}^{\pi} V_{0}^{2}cos^{4}\dfrac{\phi}{2}\left(1 - cos\phi\right)\,d\phi  \notag \\
&= \dfrac{\pi \varepsilon}{\hbar^{3}v_{f}^{2}}\dfrac{V_{0}^{2}}{4}
\label{sbsc} 
\end{flalign}
where we have used Eq.~\ref{simdos2} for density of states $ g\left(\varepsilon\right)$.

\noindent The impurities on the surface can also be of several types, each with a different density, which means that Eq.~\ref{sbsc} must be modified to reflect this situation
\begin{flalign}
\dfrac{1}{\tau} &= \sum\limits_{n_{i}}\dfrac{2 \pi}{\hbar}g\left(\varepsilon\right)\int_{0}^{\pi} n_{i}V_{0i}^{2}cos^{4}\dfrac{\phi}{2}\left(1 - cos\phi\right)\,d\phi  
\end{flalign}  
where $ n_{i} $ and $ V_{0i} $ denote a particular impurity density and the related scattering potential. 
\vspace{0.25cm}
\subsection{Scattering rate between opposite surfaces}
\vspace{0.25cm}
The two opposite surfaces in a topological insulator host states of opposite helicity; therefore, for the possibility of scattering to happen between two such states, the topological insulator film must be sufficiently thin. A thin film TI can be modeled as
\begin{equation}
H_{surf.states} = \hbar v_{f}(\sigma_{x}k_{y} - \sigma_{y}k_{x}) + \Delta_{h}\sigma_{z}
\label{hybham1}
\end{equation}
where $ 2\Delta_{h} $ is the band gap shift.
The scattering rate can be similarly computed by employing the Fermi-golden rule, the square of the matrix element now is given as $  V_{0}^{2}sin^{2}\dfrac{\phi}{2} $. Inserting the expression for matrix element in Eq.~\ref{simbz}, we get
\begin{flalign}
\dfrac{1}{\tau} &= \dfrac{2 \pi}{\hbar}g\left(\varepsilon\right)\dfrac{V_{0}^{2}\left(\hbar v_{f}k\right)^{2}}{\Delta^{2}+ \left( \hbar v_{f}k\right)^{2}}\int_{0}^{\pi}sin^{2}\dfrac{\phi}{2}cos^{2}\dfrac{\phi}{2}\left(1 - cos\phi\right)\,d\phi \notag \\
&= \dfrac{\pi \varepsilon}{4 \hbar^{3}v_{f}^{2}}\dfrac{V_{0}^{2}\left(\hbar v_{f}k\right)^{2}}{\Delta^{2}+ \left( \hbar v_{f}k\right)^{2}}
\label{sctU}
\end{flalign}

\noindent Topological insulator thin films are usually grown on substrates that render the structure asymmetric by adding an intrinsic dipole moment and breaking structural inversion symmetry (SIA). Such a thin film with an inherent SIA potential $ U $ can be described by the Hamiltonian
\begin{equation}
H_{surf.states} = \hbar v_{f}(\sigma_{x}k_{y} - \sigma_{y}k_{x}) + \Delta_{h}\sigma_{z} + U\sigma_{x}
\label{hybham2}
\end{equation}
The wave functions for the Hamiltonian in Eq.~\ref{hybham2} are
\begin{subequations}
\begin{equation}
\Psi_{\eta} = \dfrac{1}{\sqrt{2}}\begin{pmatrix}
\lambda_{\eta}(k)exp(i\theta) \\
\eta \lambda_{-\eta}(k)
\end{pmatrix}
\label{wfun3}
\end{equation}
where $ \lambda_{\eta} $ is
\begin{equation}
\lambda_{\eta} = \sqrt{1 \pm \dfrac{\Delta_{h}}{\sqrt{\left(\Delta_{h}^{2}+ \left(\hbar v_{f}k_{y}+U\right)^{2}+ \left(\hbar v_{f}k_{x}\right)^{2}\right)}}}
\label{wfun4}
\end{equation}
and $ \theta = tan^{-1}\dfrac{k_{y}}{\left(k_{x} + U\right)} $.
\end{subequations}
The corresponding eigen values are
\begin{equation}
\varepsilon\left(k\right) = \sqrt{\left(\Delta_{h}^{2}+ \left(\hbar v_{f}k_{y}+U\right)^{2}+ \left(\hbar v_{f}k_{x}\right)^{2}\right)} 
\label{evasym}
\end{equation}

\noindent The scattering time by a direct application of Fermi golden rule in the case of a carrier that tunnels through the band gap open asymmetric thin film, for instance, beginning from a $ + k $ state on the top surface and settling in to a $ -k $ state on the lower surface(opposite helicity) is 
\begin{flalign}
\dfrac{1}{\tau} &= \dfrac{2\pi}{\hbar}g\left(\varepsilon\right)V_{0}^{2}\gamma\int_{0}^{\pi}sin^{2}\dfrac{\phi}{2} cos^{2}\dfrac{\phi}{2}\left(1 - cos\phi\right)\,d\phi \notag \\
&= \dfrac{\varepsilon \pi}{4\hbar^{3}v_{f}^{2}}V_{0}^{2}\gamma
\label{gaphyb}
\end{flalign}
where $ \gamma =  \dfrac{\left(\left(\hbar v_{f}k_{y}+U\right)^{2}+\left(\hbar v_{f}k_{x}\right)^{2}\right)}{\Delta^{2}+ \left(\left(\hbar v_{f}k_{y}+U\right)^{2}+\left(\hbar v_{f}k_{x}\right)^{2}\right)} $

\vspace{0.25cm}
\subsection{Scattering due to a magnetic impurity}
\vspace{0.25cm}
Impurities impregnated on the TI surface which possess a finite magnetic moment break time reversal symmetry (TRS). A broken TRS gives a non-zero expectation value for spin polarization along an axis aligned to the outward normal to the TI surface. To compute the elastic scattering rate in this case, we must re-evaluate the spin suppression factor $ \zeta\left(s,s^{'}\right)$ which was set to $ cos^{2}\dfrac{\phi}{2} $. The spin polarization vectors, bearing in mind the helical structure of the surface states are therefore given by
\begin{subequations}
\begin{eqnarray}
\langle S_{x}\rangle = \dfrac{\hbar}{2}\langle \Psi_{+}\vert \begin{pmatrix}
0 & 1 \\ 1 & 0
\end{pmatrix}\vert \Psi_{+} \vert \notag \\
= \dfrac{\hbar v_{f}k}{\sqrt{\Delta^{2}+ \left(\hbar v_{f}k\right)^{2}}}cos\theta_{1}
\end{eqnarray}
\begin{eqnarray}
\langle S_{y}\rangle = \dfrac{\hbar}{2}\langle \Psi_{+}\vert \begin{pmatrix}
0 & -i \\ i & 0
\end{pmatrix}\vert \Psi_{+} \vert \notag \\
= \dfrac{\hbar v_{f}k}{\sqrt{\Delta^{2}+ \left(\hbar v_{f}k\right)^{2}}}sin\theta_{1}
\end{eqnarray}
\begin{eqnarray}
\langle S_{z}\rangle = \dfrac{\hbar}{2}\langle \Psi_{+}\vert \begin{pmatrix}
1 & 0 \\ 0 & -1
\end{pmatrix}\vert \Psi_{+} \vert \notag \\
= \dfrac{\Delta}{\sqrt{\Delta^{2}+ \left(\hbar v_{f}k\right)^{2}}}
\end{eqnarray}
\end{subequations}
The final spin polarization vector is  
\begin{equation}
\textbf{S} = S_{x}\widehat{i} + S_{y}\widehat{j} + S_{z}\widehat{k}
\end{equation}
The spin suppression factor between two states $\vert k\left(\theta_{1}\right),s\rangle $ and $\vert k^{'}\left(\theta_{2}\right),s^{'}\rangle $  suffering an angular elastic $ \left(\vert k \vert = \vert k^{'} \vert \right)$ scattering of $ \phi = \theta_{1}-\theta_{2}  $ is therefore
\begin{subequations}
\begin{flalign}
\zeta\left(s,s^{'}\right) &= cos^{2}\left[\dfrac{1}{2}cos^{-1}\dfrac{s \cdot s^{'}}{\vert s \vert \vert s^{'} \vert}\right] \notag \\
&= cos^{2}\left[\dfrac{1}{2}cos^{-1}\dfrac{\Omega_{1}^{2}cos\phi + \Omega_{2}^{2}}{\Omega_{1}^{2}+ \Omega_{2}^{2}}\right]
\label{spsc}  
\end{flalign}
where
\begin{eqnarray}
\Omega_{1} = \dfrac{\hbar v_{f}k}{\sqrt{\Delta^{2}+ \left(\hbar v_{f}k\right)^{2}}} \notag \\
\Omega_{2} = \dfrac{\Delta}{\sqrt{\Delta^{2}+ \left(\hbar v_{f}k\right)^{2}}}
\label{omegas}
\end{eqnarray}
The magnetic field induced band gap is $ \Delta = \dfrac{1}{2}g\mu_{B}B_{z}$.
\end{subequations}

\noindent As before, the scattering time can be computed using Eq.~\ref{simbz}; further, in presence of a magnetic field, time reversal symmetry is lost and back scattering is no longer forbidden. Carriers on the surface after scattering can therefore occupy states of opposite helicity or in other words with a reversed $ k $ vector (for example, $ +k_{1} $ to $ -k_{2}$, such that $ \vert k_{1}\vert = \vert -k_{2}\vert$ for an elastic scattering). The scattering times when the final scattered states have identical ($ +k_{1} $ to $ + k_{2}$ and $\vert k_{1}\vert = \vert k_{2}\vert$) or opposite helicity will be different and can be calculated as follows. We first establish the scattering rate between states of identical helicity.
\begin{subequations}
\begin{equation}
\dfrac{1}{\tau} = \dfrac{2 \pi}{\hbar}g\left(\varepsilon\right)\int_{0}^{\pi}d\phi\vert\langle\Psi_{+}\vert V_{0}\zeta\left(s,s^{'}\right)\vert \Psi_{+}\rangle\vert^{2}\left(1 -cos\phi\right) 
\end{equation}
Inserting the spin scattering factor using Eq.~\ref{spsc} and wave functions from Eq.~\ref{wfun1} and carrying out the straight forward algebra yields
\begin{equation}
\dfrac{1}{\tau} = \dfrac{2\varepsilon}{\hbar^{3}v_{f}^{2}}V_{0}^{2}\int_{0}^{\pi}d\phi cos^{2}\vartheta\left\lbrace cos^{2}\dfrac{\phi}{2} + \kappa sin^{2}\dfrac{\phi}{2} \right\rbrace\left(1 - cos\phi\right)
\end{equation}
where $ \kappa = \dfrac{\Delta^{2}}{\Delta^{2}+ \left(\hbar v_{f}k\right)^{2}} $
The equation can be numerically evaluated as shown in Sec III. A similar integral for scattering between states of opposite helicity can be written
\begin{equation}
\dfrac{1}{\tau} = \dfrac{2\varepsilon}{\hbar^{3}v_{f}^{2}}V_{0}^{2}\int_{0}^{\pi}d\phi cos^{2}\vartheta\left\lbrace \dfrac{\left(\hbar v_{f}k\right)^{2}}{\Delta^{2}+ \left(\hbar v_{f}k\right)^{2}}sin^{2}\dfrac{\phi}{2}\right\rbrace\left(1 - cos\phi\right)
\end{equation}
where $ \vartheta = \left[\dfrac{1}{2}cos^{-1}\dfrac{\Omega_{1}^{2}cos\phi + \Omega_{2}^{2}}{\Omega_{1}^{2}+ \Omega_{2}^{2}}\right] $ and $ \Omega_{1} $ and $ \Omega_{2} $ are defined in Eq.~\ref{omegas}.
\end{subequations}

\vspace{0.25cm}
\section{Influence of warping on surface conductivity}
\vspace{0.25cm}

Under a weak external force, the deviation of the electron distribution from the thermal equilibrium value can be assumed to be small which allows us to linearize Boltzmann distribution within the relaxation time approximation. The conductivity can therefore be written as
\begin{equation}
\sigma = \dfrac{e^{2}v_{f}^{2}}{2}\int dE\, g\left(E\right)\tau\left(-\dfrac{\partial f}{\partial E}\right) 
\label{econd}   
\end{equation}
which at $ T = 0 $ yields $ \sigma = \dfrac{e^{2}v_{f}^{2}}{2}g\left(E\right)\tau $, where we substitute for the diffusion constant $ D $ in the original Einstein relation $ \sigma = e^{2}g\left(E\right)D $ as $ D = \dfrac{v_{f}^{2}\tau}{2} $ for a two-dimensional system. The scattering time for each scenario considered can be inserted from their respective expressions derived above. Finally, we note that the expressions derived for scattering time and conductivity undergo a modification if warping terms are explicitly included in the analysis. More precisely, the band gap $ \Delta $ modifies to $ \Delta + \hbar^{3}\lambda k^{3}cos3\theta $ and the density of states also undergoes a change as shown below
\begin{subequations}
\begin{flalign}
D(\epsilon)_{warp} &= \int \dfrac{dk}{\left(2\pi\right)^{2}}\delta\left(\epsilon - E(k)\right) \\
&= \dfrac{1}{\left(2\pi\right)^{2}}\int_{0}^{2\pi} d\theta \int_{0}^{\infty}k\,dk \sum\limits_{j}\dfrac{\delta(k-k_{j})}{\vert g^\prime(k_{j})\vert} \\
&= \dfrac{1}{\left(2\pi\right)^{2}}\sum\limits_{j}\int_{0}^{2\pi} d\theta \dfrac{k_{j}}{\vert g^\prime(k_{j})\vert}
\label{finaldos} 
\end{flalign}
\end{subequations}
We have used the identity $ \delta\left(g(x)\right) = \sum\limits_{j}\dfrac{\delta(x-x_{j})}{\vert g^\prime(x_{j})\vert}$ such that $ g(x_{j}) = 0 $ and there are no multiple zeros. $ x_{j} $ is a simple zero of $ g(x) $. 
The function $ g(x) $ takes the form
\begin{equation}
g(x) = \epsilon - \sqrt{\hbar v_{f}^{2}k^{2} + \lambda^{2}\hbar^{6}k^{6}cos^{2}(3\theta)} 
\label{zeroseq}
\end{equation}
The band gap split $ \Delta $ is ignored since it is swamped by the warping correction at points in momentum space away from the Dirac cones at $ \Gamma $.
The $ k_{j} $ in Eq.~\ref{finaldos} are obtained by determining roots of Eq.~\ref{zeroseq}. Expanding Eq.~\ref{zeroseq} yields
\begin{subequations}
\begin{equation}
\hbar v_{f}^{2}k^{2} + \lambda^{2}\hbar^{6}k^{6}cos^{2}(3\theta) - \epsilon^{2} = 0
\label{zeroseqsq}
\end{equation} 
Rearranging, Eq.~\ref{zeroseqsq} is written as a cubic equation with $ k^{2} $ as the variable
\begin{equation}
\lambda^{2}\hbar^{6}cos^{2}(3\theta)\left(k^{2}\right)^{3} + \hbar v_{f}^{2}k^{2} - \epsilon^{2} = 0
\label{warpeq}
\end{equation}
\end{subequations}
The real solution for $ k^{2} $ is of the form
\begin{equation} 
\left(\omega_{1} + \omega_{2}\right)^{1/3} - \left(\vert\omega_{1} - \omega_{2}\right\vert)^{1/3}  
\label{ksol}
\end{equation}
where $ \omega_{1} $ and $ \omega_{2} $ are defined as follows
\begin{subequations}
\begin{equation}
\omega_{1} = \dfrac{1}{2}\left(\dfrac{\epsilon}{\lambda\hbar^{3}cos(3\theta)}\right)^{2}
\label{wr1}
\end{equation}
and
\begin{equation}
\omega_{2} = \sqrt{\dfrac{1}{27}\left(\dfrac{\hbar v_{f}}{\lambda\hbar^{3}cos(3\theta)}\right)^{6}  + \dfrac{1}{4}\left(\dfrac{\epsilon}{\lambda\hbar^{3}cos(3\theta)}\right)^{4}} 
\label{wr2}
\end{equation}
\end{subequations}
Finally evaluating the derivative $ g^{\prime}(x)$, we have
\begin{equation}
g^{\prime}(x)= \mp \dfrac{\hbar v_{f}^{2}k + 3\lambda^{2}\hbar^{6}k^{5}cos^{2}\left(3\theta\right)}{\sqrt{\hbar v_{f}^{2}k^{2} + \lambda^{2}\hbar^{6}k^{6}cos^{2}\left(3\theta\right)}}
\label{gderv}
\end{equation}
Putting all of them together, the density of states at any given zero of $ g(x) $, $ x_{j} $ can be written as
\begin{subequations}
\begin{equation}
D(\epsilon) = \dfrac{1}{\left(2\pi\right)^{2}}\int_{0}^{2\pi} d\theta \ k\dfrac{\sqrt{\left(\hbar v_{f}\right)^{2}k^{2} + \lambda^{2}\hbar^{6}k^{6}cos^{2}\left(3\theta\right)}}{\left(\hbar v_{f}\right)^{2}k + 3\lambda^{2}\hbar^{6}k^{5}cos^{2}\left(3\theta\right)}
\end{equation}
Simplifying, we get
\begin{equation}
D(\epsilon) = \dfrac{1}{\left(2\pi\right)^{2}}\int_{0}^{2\pi} d\theta\ k \dfrac{\epsilon}{\left(\hbar v_{f}\right)^{2}k + 3\lambda^{2}\hbar^{6}k^{5}cos^{2}\left(3\theta\right)}
\end{equation}
Noting that energy $ \epsilon $ is a constant, the final form for $ D(\epsilon) $ is 
\begin{equation}
D(\epsilon) = \dfrac{\epsilon}{\left(2\pi\right)^{2}}\int_{0}^{2\pi} d\theta \dfrac{k^{2}}{\epsilon^{2} + 2\lambda^{2}\hbar^{6}k^{6}cos^{2}\left(3\theta\right)}
\label{simdos}
\end{equation}
In simplifying the above expression, we have used the Dirac cone energy expression $ \sqrt{\hbar v_{f}^{2}k^{2} + \lambda^{2}\hbar^{6}k^{6}cos^{2}(3\theta)} $. 
Inserting for $ k $ from Eq.~\ref{ksol},~\ref{wr1}, and ~\ref{wr2}, the final expression for density of states (Eq.~\ref{simdos}) is evaluated numerically in Section IV.
\end{subequations}

\vspace{0.5cm}
\section{Results}
\vspace{0.5cm}

All results derived in this paper use the two-band model (Eq.~\ref{dss}) which employs phenomenological parameters to describe band gap splitting and inter-surface coupling in case of an asymmetric thin film. These parameters can be directly established from a first-principles calculation~\cite{zhang2010first,park2013ageing} or obtained, as done here, by numerically diagonalizing the four-band Hamiltonian of Eq.~\ref{eqn1}. The required parameters are read off the dispersion plot. The Fermi velocity of the surface states is taken as 5 $ \times 10^{5} $ m/s for all calculations shown.
\vspace{0.25cm}
\subsection{Non-magnetic impurities}
\begin{figure}[h]
\includegraphics[scale= 1]{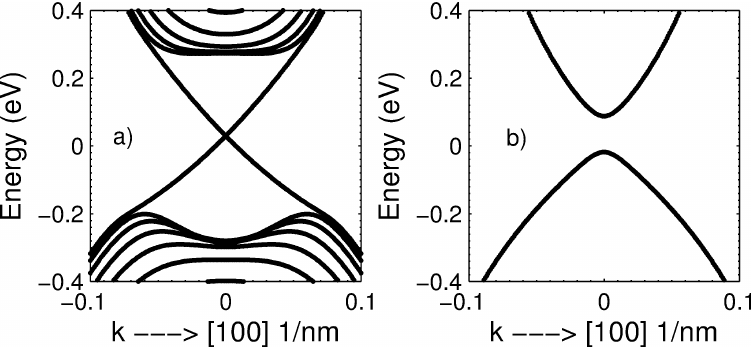}
\caption{Dispersion of a 9.0 $ \mathrm{nm} $ thick Bi$_{2}$Se$_{3}$ topological insulator slab around the Dirac point (Fig.~\ref{fig1}a) at 0.02 $\mathrm{eV}$. The dispersion of the thin film (3.0 $ \mathrm{nm} $ thick) (Fig.~\ref{fig1}b) shows two Dirac hyperbolas when the surface states hybridize.}
\label{fig1}
\end{figure}

\vspace{0.25cm}
We first plot the dispersion relationships (Fig.~\ref{fig1}) for a nine and three quintuple layer Bi$_{2}$Se$_{3}$ topological insulator film. Each quintuple layer is about 1.0 $ \mathrm{nm}$. The three quintuple layer film has its top and bottom surfaces coupled (hybridized surfaces) which induces a finite band gap. On the surface of a  9.0 $ \mathrm{nm} $ thick Bi$_{2}$Se$_{3}$ topological insulator slab which has zero-gap states, a uniform, dilute, and non-interacting spherical impurity scattering potential of 10.0 $ \mathrm{meV} $ is assumed to be present. Electrons scatter from such fixed impurities with no internal excitations elastically. The scattering time is computed at an energy equal to the Fermi level which is set to 100.0 $ \mathrm{meV} $. At this energy, the density of states (DOS), using Eq.~\ref{simdos1} is 2.93 $\times 10^{-6} \mathrm{meV^{-1}A^{-2}}$. Inserting the DOS in the scattering time equation (Eq.~\ref{sbsc}) yields a value equal to 9.076 $\times 10^{-10} \mathrm{seconds}$. It is important to note that the Fermi-level can be raised up to 0.2 $ \mathrm{eV} $, beyond which we reach the bulk conduction bands of Bi$_{2}$Se$_{3}$. The bulk conduction bands of the slab can be seen to begin from the 0.2 $ \mathrm{eV} $ mark in Fig.~\ref{fig1}.

\begin{figure}
\includegraphics[scale=0.7]{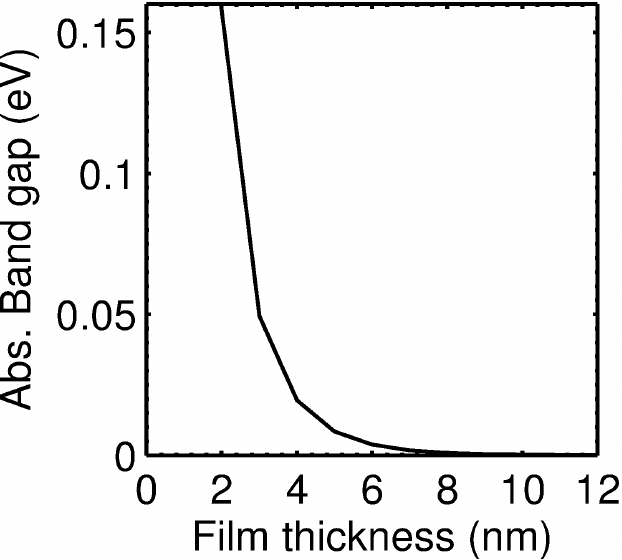}
\caption{Band-gap opening as a function of Bi$_{2}$Se$_{3}$ film thickness. A band gap opens because the wave function from the two surfaces penetrate the bulk and hybridize.}
\label{fig2}
\end{figure}

For the case of a thin TI film, the possibility of an electron tunneling through the bulk and reaching a state of opposite helicity on the second surface must be accounted for. The variation in band gap of the surface states for a thin film with varying thickness is plotted in Fig.~\ref{fig2}.
Retaining the parameters from the case of scattering between states of identical helicity and setting the band gap $ \Delta $ in Eq.~\ref{sctU} to 100.0 $ \mathrm{meV} $, the corresponding scattering time is determined to be 8.309 $\times 10^{-10} \mathrm{seconds}$. The scattering time is similar to the one obtained when scattered carriers are constrained only to occupy another state of identical helicity on the same surface. It is easy to see from Eq.~\ref{sbsc} and Eq.~\ref{sctU} that the ratio of scattering times between the two cases is  $ \dfrac{\left(\hbar v_{f}k\right)^{2}}{\Delta^{2}+ \left( \hbar v_{f}k\right)^{2}} $. The ratio is significant when $ k $ is close to the $ \Gamma $ point or the Dirac cone but tends to unity at larger $ k $ values since the hybridization induced band gap split is generally a small number. As pointed out later, at points far away from $ \Gamma $ where the Dirac crossing happens, the density of states using Eq.~\ref{simdos2} is inadequate. 

\begin{figure}
\includegraphics[scale=1.15]{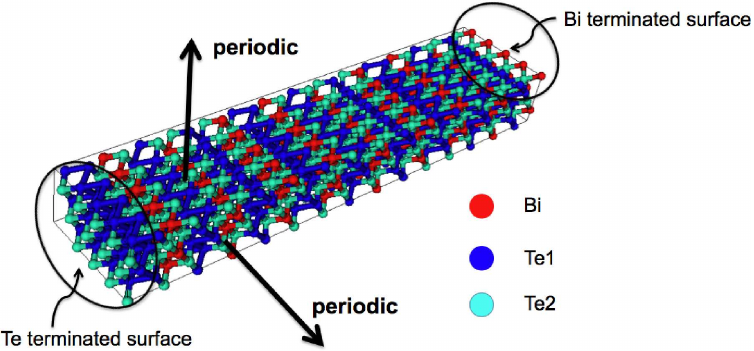}
\caption{A 9.0 nm thick Bi$_{2}$Te$_{3}$ ultra-thin body oriented along the x-axis.The two surfaces have Bi and Te termination thus making them chemically non-equivalent.}
\label{fig3}
\end{figure}

\begin{figure}
\includegraphics[scale=1]{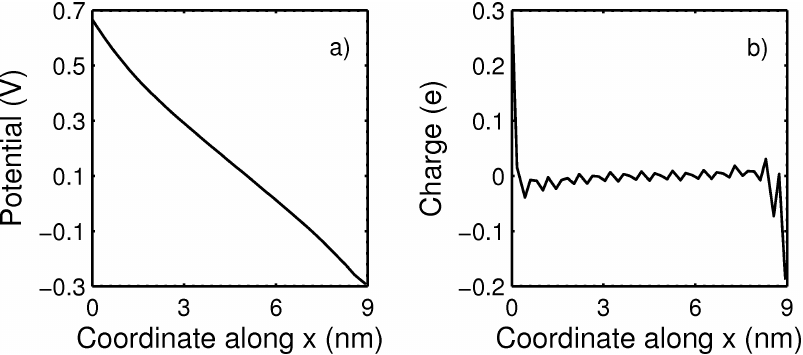}
\caption{The spatially-dependent electrostatic potential (Fig.~\ref{fig4}a) and charge on each atomic node is plotted against the \textit{x} coordinate of the Bi$_{2}%
$Te$_{3}$ thin film. This thin-film has Bi and Te termination on the surfaces. An oscillating charge pattern (Fig.~\ref{fig4}b) is obtained for non-equivalent
surface termination.}
\label{fig4}
\end{figure}

We next turn our attention to a TI thin film grown on a substrate that renders the structure asymmetric by creating a spatially-dependent dipole moment. A simple way to mimic the effect of a substrate is to model the film with two different surface terminations as depicted in Fig.~\ref{fig3}. The asymmetric potential gives structural inversion asymmetry (SIA) which means that the two surfaces do not have equi-energetic Dirac cones. As shown in Fig.~\ref{fig4}, a TI film with bismuth and tellurium surface termination possesses a non-zero dipole moment~\cite{psthe,n5,fonseca2013efficient} with an oscillating charge pattern in bulk of the film.
In absence of the SIA potential, the surface states consist of spin degenerate conduction and valence bands~\cite{sengupta2015proximity} separated by the  hybridization gap $ \Delta $ while the substrate induced asymmetry leads to a Rashba splitting.~\cite{dedkov2008rashba} The actual band gap (Eq.~\ref{evasym}) varies with the overall dipole moment. For the thin film considered here, $ U $ is found to be 1.0 $ \mathrm{eV} $~\cite{psthe} and using identical parameters as before, the scattering time is 9.025 $\times 10^{-10} \mathrm{seconds}$. In evaluating the scattering time, the wave vector is assumed to only have a $ k_{x} $ component equal to 0.1 1/\AA. 

The scattering time in all the three cases is roughly the same which suggests that the band gap opening and the dipole moment do not significantly affect the results since the factors  $ \dfrac{\left(\hbar v_{f}k\right)^{2}}{\Delta^{2}+ \left( \hbar v_{f}k\right)^{2}} $ and  $ \dfrac{\left(\left(\hbar v_{f}k_{y}+U\right)^{2}+\left(\hbar v_{f}k_{x}\right)^{2}\right)}{\Delta^{2}+ \left(\left(\hbar v_{f}k_{y}+U\right)^{2}+\left(\hbar v_{f}k_{x}\right)^{2}\right)} $ from Eq.~\ref{sctU} and Eq.~\ref{gaphyb} at moderate $ k $ vectors are numerically close. These factors, nonetheless, have a relatively important contribution when the ratio of band splitting energy ($\Delta) $ to the Fermi-energy $ \hbar v_{f} k $ is not negligible which is possible for very small values of the momentum vector.

All the scattering times discussed here include the spin suppression factor whose contribution can be gauged by evaluating the scattering rate equation for a zero-gap TI with impurities (Eq.~\ref{sbsc}) without the $ cos^{2}\dfrac{\phi}{2} $ term. Such a calculation gives a scattering time of 2.8 $\times 10^{-10} \mathrm{seconds}$, which is a reduction by a factor of 3.24. A shorter scattering time signifies lower mobility and conductivity~\cite{ferry2009transport} which can be easily explained by noting that the high mobility of surface states is attributed to spin-protected forbidden back scattering. The condition to forbid back scattering is relaxed by ignoring the spin-suppression factor of $ cos^{2}\dfrac{\phi}{2} $. We also note, as shown in Ref.~\onlinecite{sengupta2015evaluation}, the scattering time in a gapped topological insulator(trivial case) is  four-fold higher compared to a zero-gapped topological insulator (non-trivial case). The factor of 3.24 in the present case where the suppression factor controls back scattering matches well with the previously derived result.
In passing we mention here, that the dipole moment, in principle, at a critical value, can close the band gap and re-open as an instance of topological phase transition.~\cite{li2010chern}.

\vspace{0.25cm}
\subsection{Magnetic impurities and warped bands}
\vspace{0.25cm}
The helical nature~\cite{yazyev2010spin,wray2011topological} of the surface states of a pristine topological insulator ensures that there is a definite in-plane spin polarization. Magnetic impurities on the surface give an additional out-of-plane spin component~\cite{liu2009magnetic} thus essentially turning the spin-polarization to a three-dimensional vector. Scattering times corresponding to a 9.0 $ \mathrm{nm} $ and 3.0 $ \mathrm{nm} $ thin film are presented for discussion. For the 3.0 $ \mathrm{nm} $ thin film, we consider the case of a carrier tunneling to the other surface and occupying a state of opposite helicity. Proceeding along same lines as for non-magnetic impurities with the spin suppression factor now modified such that spins are aligned to the effective spin-polarization vector, yields scattering times equal to 7.530 $\times 10^{-10} \mathrm{seconds}$ and 6.935 $\times 10^{-10} \mathrm{seconds}$ for the 9.0 $ \mathrm{nm} $ and 3.0 $ \mathrm{nm} $ TI films respectively. As a first check, the scattering times are shorter than 9.076 $\times 10^{-10} \mathrm{seconds}$, which is the corresponding result for a zero-gap topological insulator film. The presence of a band gap reduces the scattering time, augments the large angle scattering and consequently lowers the mobility as expected. In calculations shown here, we have explicitly neglected the orbital coupling of the magnetic field through a Peierls substitution and retained only the spin coupling through the Zeeman term. The Zeeman term can be attributed to an externally applied magnetic field or the exchange field of a ferromagnet layered on the surface of a TI.~\cite{zhang2012interplay} 

\begin{figure}
\includegraphics[scale=0.8]{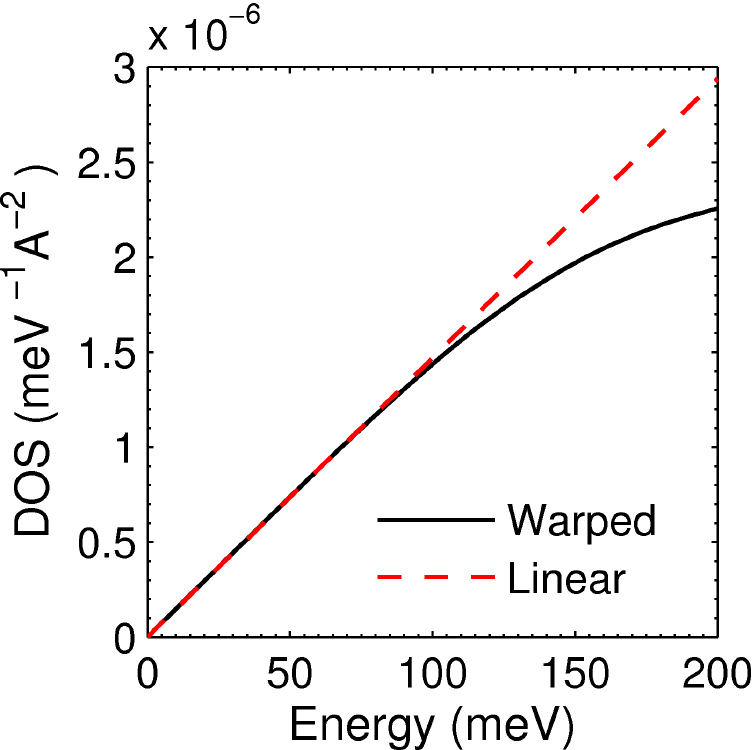}
\caption{The density of states for a warped dispersion deviates from the expression obtained for linear bands at higher energies. At higher energies, the warping terms significantly contribute to the overall energy spectrum of the TI surface states. The warping strength in this calculation was set to 500.0 $ \mathrm{ev \AA^{-3}}$.}
\label{fig5}
\end{figure}

At this point, it is worthwhile to establish the change in scattering rates due to warped nature of bands, if the momentum vectors are not in the immediate proximity of the $ \Gamma $ point. The warped bands modify the density of states as derived in Eq.~\ref{simdos}. A comparative plot of density of states obtained from a linear dispersion and with warping terms included is shown in Fig.~\ref{fig5}. The density of states for the warped case was determined by numerically integrating Eq.~\ref{simdos}.

\begin{figure}
\includegraphics[scale=0.8]{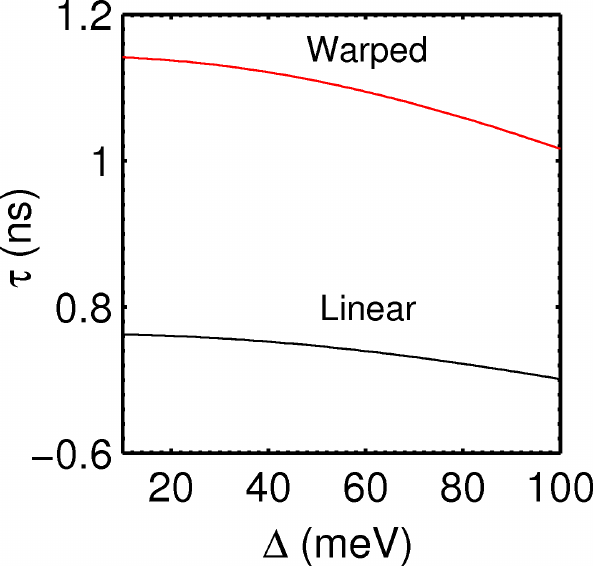}
\caption{The scattering time on surface of a magnetically gapped topological insulator is higher for the warped band model (warping strength = 500.0 $ \mathrm{ev \AA^{-3}}$ ) because of a lower density of states. The scattering time also decreases with a greater magnetic field induced band gap indicating lower mobility or a higher magnetoresistance.}
\label{fig6}
\end{figure}

The scattering time for a carrier described by a warped Hamiltonian (Eq.~\ref{warping}) on surface of a zero-gap topological insulator whose initial and final states are of same helicity is 1.541 $\times 10^{-9} \mathrm{seconds}$. The scattering time is higher compared to 7.530 $\times 10^{-10} \mathrm{seconds}$ because of a lower density of states with warped bands. It is well-known from semi-classical transport theory~\cite{ziman1972principles} that a lower density of states reduces scattering and increases mobility. Scattering times for a range of band gaps for the linear and warped dispersion models is shown in Fig.~\ref{fig6}. The scattering time decreases almost linearly under  higher magnetic fields(increased band gaps) for both the Dirac and warped dispersion which is indicative of a greater magnetoresistance. The origin of the enhanced magnetoresistance is through a large angle scattering $ \left(1 - cos\theta\right) $ term that enters the analysis in the scattering rate equation. Further, since the electric conductivity, as evident from Eq.~\ref{econd}, is directly proportional to the scattering time, the warped model would yield a higher value. In other words, electric conductivity is an implicit function of momentum space; at $ k $ points away from $\Gamma $, the warped Hamiltonian which is a better description of electronic spectrum of surface states yields lower density of states and a higher electron velocity.

\vspace{0.25cm}
\section{Surface conductivity and mean free path}
\vspace{0.25cm}
The conductivity on the impure surface of a topological insulator can be estimated by a direct application of Eq.~\ref{econd}. We consider a non-magnetic impurity of density $ n_{i} = 0.5 \times 10^{10} $. The scattering potential is 500.0 $ \mathrm{eV} $. The scattering event takes place at a Fermi level set to 200.0 $ \mathrm{meV} $ for which the density of states (using Eq.~\ref{simdos1} for linear bands) is 5.877 $\times 10^{-6} \mathrm{meV^{-1}A^{-2}}$. The scattering time for this case turns out to be 9.076 $ \times 10^{-14} $ seconds. Putting all these numbers in Eq.~\ref{econd}, the surface conductivity, expressed in terms of the quantum conductance $ \sigma = \dfrac{2e^{2}}{h}$ is 137.595$\sigma$. The conductivity at large $ k $ values is reduced since the corresponding density of states must now be computed with warped bands. The DOS with warped bands at 200.0 $ \mathrm{meV} $ evaluates to (by numerical integration of Eq.~\ref{simdos}) 4.51$\times 10^{-6} \mathrm{meV^{-1}A^{-2}}$ while the scattering time is 1.183 $ \times 10^{-13} $ seconds. The surface conductivity value of 137.721$ \sigma $ is only marginally higher than the corresponding 137.595 $ \sigma $ for linear bands. Another useful that can be easily computed is the mean free path given as $ v_{f}\tau $. The numbers for linear and warped bands are 45.38 $ \mathrm{nm} $ and 59.15 $ \mathrm{nm} $ respectively.

While the above analysis does not account for scattering events which involve phonons and charged impurities, the mean free paths obtained suggest that transport in a topological insulator channel material for a conventional field-effect transistor in a miniaturized device will be ballistic. The experimentally determined mean free path in GaAs, a high-mobility semiconductor, is approximately 34.0 $ \mathrm{nm} $ at room temperature.

\vspace{0.25cm}
\section{Conclusion}
\vspace{0.25cm}
In this work we have theoretically presented the evaluation of scattering times on the surface of a 3D topological insulator using Boltzmann transport. The role of spin suppression factor which comes from the topological nature of the states and explains the higher mobility on a TI surface is highlighted. It is found that in case of a magnetically split TI surface, the alignment of the spin polarization vectors to the effective magnetic field reduces the overall scattering time. The influence of density of states from a warped model is analyzed and a higher mobility and electric conductivity value is established. The enhancement is attributed to the reduced density of states. In carrying out these calculations we have assumed a uniform Fermi velocity for the surface states though that may not be true for all materials that exhibit topological insulator behaviour. In particular, the topological Kondo insulator~\cite{neupane2013surface,dzero2012theory} which has Fermi pockets~\cite{zhu2013polarity} with varying velocities is a case in point. Also, in a more careful calculation, which could be potential future work, preferential scattering directions dependent on the underlying crystal symmetry must be considered. Finally, we note that because of the helical structure of the surface state, the spin polarization vector must have a definite polarization for a given $ k $ state. It is tacitly assumed that the spin relaxation happens at a much faster rate than momentum relaxation; in case the spin-relaxation time, through the well-known Dyakanov-Perel or Elliot-Yafet relaxation mechanisms~\cite{huertas2009spin,dyakonov2008spin} is comparable to the momentum relaxation time, the overall scattering rate and mobility will be altered. This aspect has not been investigated here.

\vspace{0.4cm}
\begin{acknowledgements}
\vspace{0.4cm}
This work at Boston University was supported in part by the BU Photonics Center and U. S. Army Research Laboratory through the Collaborative Research Alliance (CRA) for MultiScale multidisciplinary Modeling of Electronic materials (MSME). PS also thanks Saumitra Mehrotra of Freescale Semiconductors for reading the manuscript.
\end{acknowledgements} 

\bibliographystyle{apsrev}
\bibliography{References} 
\end{document}